\documentclass[12pt]{article}
\usepackage{amsmath,amssymb,epsfig}
\makeatletter \@addtoreset{equation}{section} \makeatother
\renewcommand{\theequation}{\thesection.\arabic{equation}}
\newcommand{\ba}{\begin{array}}
\newcommand{\ea}{\end{array}}
\newcommand{\beq}{\begin{equation}}
\newcommand{\eeq}{\end{equation}}
\newcommand{\bea}{\begin{eqnarray}}
\newcommand{\eea}{\end{eqnarray}}

\def\bce{\begin{center}}
\def\ece{\end{center}}

\def\nonu{\nonumber}

\def\pa{\partial}

\def\be{\beta}

\def\de{\delta}

\def\ep{\epsilon}

\def\diag{\mathop{\rm diag}}

\def\eps6{{\displaystyle \mathop{\epsilon}^{6}}{}}

\def\nab6{{\displaystyle \mathop{\nabla}^{6}}{}}

\def\0{{\sst{(0)}}}
\def\1{{\sst{(1)}}}
\def\2{{\sst{(2)}}}
\def\3{{\sst{(3)}}}
\def\4{{\sst{(4)}}}
\def\5{{\sst{(5)}}}
\def\6{{\sst{(6)}}}
\def\7{{\sst{(7)}}}
\def\8{{\sst{(8)}}}

\def\ba{\begin{array}}
\def\ea{\end{array}}
\def\beq{\begin{equation}}
\def\eeq{\end{equation}}
\def\be{\begin{equation}}
\def\ee{\end{equation}}

\def\diag{\mathop{\rm diag}}

\def\eps{\epsilon}

\def\ba{\begin{array}}
\def\ea{\end{array}}
\def\beq{\begin{equation}}
\def\eeq{\end{equation}}
\def\be{\begin{equation}}
\def\ee{\end{equation}}

\def\diag{\mathop{\rm diag}}

\def\eps{\epsilon}

\def\eps6{{\displaystyle \mathop{\epsilon}^{6}}{}}

\def\nab6{{\displaystyle \mathop{\nabla}^{6}}{}}

\newcommand{\bean}{\begin{eqnarray*}}
\newcommand{\eean}{\end{eqnarray*}}

\begin{document}
\thispagestyle{empty} \addtocounter{page}{-1}
   \begin{flushright}
\end{flushright}

\vspace*{1.3cm}
  
\centerline{ \Large \bf  New ${\cal N}=2$ Supersymmetric Membrane Flow }
\vspace{.3cm} 
\centerline{ \Large \bf   In Eleven-Dimensional Supergravity}
\vspace*{1.5cm}
\centerline{{\bf Changhyun Ahn }
} 
\vspace*{1.0cm} 
\centerline{\it  
Department of Physics, Kyungpook National University, Taegu
702-701, Korea} 
\vspace*{0.8cm} 
\centerline{\tt ahn@knu.ac.kr 
} 
\vskip2cm

\centerline{\bf Abstract}
\vspace*{0.5cm}

We construct the 11-dimensional lift of 
the known ${\cal N}=2$ supersymmetric RG flow solution in 4-dimensional
${\cal N}=8$ gauged supergravity.  The squashed and stretched 
7-dimensional internal metric 
preserving $SU(2) \times U(1) \times U(1)_R$ symmetry contains 
an Einstein-Kahler 2-fold which is a
base manifold of 5-dimensional Sasaki-Einstein $Y^{p, q}$ space found
in 2004.   
The nontrivial $r$(transverse to the domain wall)-dependence 
of the $AdS_4$ supergravity fields makes 
the Einstein-Maxwell equations consistent not only at the critical
points but also along the supersymmetric whole RG flow connecting two
critical points.  
With an appropriate 3-form gauge field, 
we find an exact solution to the 11-dimensional Einstein-Maxwell
equations corresponding to the above lift of the $SU(2) \times
U(1) \times U(1)_R$-invariant RG flow. 
The particular limits of this solution give rise to
the previous solutions with $SU(3) \times U(1)_R$ or
$SU(2) \times SU(2) \times U(1)_R$.

\baselineskip=18pt
\newpage
\renewcommand{\theequation}
{\arabic{section}\mbox{.}\arabic{equation}}

\section{Introduction}

The
${\cal N}=6$ $U(N) \times U(N)$ 
Chern-Simons matter theory
with level $k$ in 3-dimensions
is described as the low energy limit of $N$ M2-branes at 
${\bf C}^4/{\bf Z}_k$ singularity \cite{ABJM}.
When $k=1, 2$, the full ${\cal N}=8$ supersymmetry
is preserved while for  $k > 2$,
the supersymmetry is broken to the ${\cal N}=6$ supersymmetry. 
The matter contents and the superpotential  of this theory
are the same as for the D3-branes on the conifold \cite{KW}. 
The RG flow
between the UV point and 
the IR point of the 3-dimensional 
gauge theory can be determined from the gauged ${\cal N}=8$ 
supergravity in 4-dimensions via AdS/CFT 
correspondence \cite{Maldacena}. 
The holographic supersymmetric
RG flow equation connecting ${\cal N}=8$ $SO(8)$ point 
to ${\cal N}=2$ $SU(3) \times U(1)$ point has been studied in 
\cite{AP,AW} where the $U(1)$ symmetry here can be identified with $U(1)_R$
symmetry of 3-dimensional theory coming from the ${\cal N}=2$ 
supersymmetry while
those from ${\cal N}=8$ $SO(8)$ point 
to
${\cal N}=1$ $G_2$ point has been studied in 
\cite{AW,AI}.
The 11-dimensional M-theory lifts of these 
RG flow equations have been found in \cite{CPW,AI} by solving the
Einstein-Maxwell equations in 11-dimensions with nonzero field
strengths in the internal space.

The mass deformed $U(2) \times U(2)$
Chern-Simons matter theory with level $k=1, 2$ 
preserving global $SU(3) \times U(1)_R$ symmetry has been studied 
in \cite{Ahn0806n2,BKKS} while
the mass deformation for this theory preserving $G_2$
symmetry  has been described and 
the nonsupersymmetric 
RG flow equations preserving $SO(7)^{\pm}$ symmetries 
have been discussed in \cite{Ahn0806n1}.  
The holographic
RG flow equations connecting ${\cal N}=1$ $G_2$ point 
to ${\cal N}=2$ $SU(3) \times U(1)_R$ point have been 
found in \cite{BHPW}.  
Moreover, the ${\cal N}=4$ and ${\cal N}=8$ RG flows have been 
studied and  
further developments on  
the gauged ${\cal N}=8$ supergravity in four-dimensions have been done
in \cite{AW09}. 
Recently, the spin-2 Kaluza-Klein modes around 
a warped product of $AdS_4$ and a seven-ellipsoid which has 
global $G_2$ symmetry are discussed in \cite{AW0907}. 
Furthermore, the gauge dual with the symmetry of
$SU(2) \times SU(2) \times U(1)_R$ for the second 11-dimensional lift of 
${\cal N}=2$ $SU(3) \times U(1)_R$-invariant solution 
in 4-dimensional supergravity 
is described in \cite{AW0908}.

The seven-sphere ${\bf S}^7$ in the internal space 
can be realized by ${\bf S}^1$-fibration
over ${\bf CP}^3$ \cite{NP} space where the standard Fubini-Study
metric on the ${\bf CP}^3$ space has ${\bf CP}^2$
space \cite{PW,CPW} 
or ${\bf CP}^1
\times {\bf CP}^1$ space \cite{CPW,CLP}.
In particular, the $U(1)$ bundle over 
${\bf CP}^1 \times {\bf CP}^1$ space is known as 5-dimensional
Sasaki-Einstein $T^{1,1}$ space \cite{Cd}.    
In \cite{CPW}, they have found two different 11-dimensional solutions
where the first has ${\bf CP}^2$ space with $SU(3) \times U(1)_R$
symmetry
and the second has ${\bf CP}^1
\times {\bf CP}^1$ space with $SU(2) \times SU(2) \times U(1)_R$ symmetry. 
Note that the Ricci tensor for the first
solution with frame basis is exactly the same as the one of the second 
solution, by assuming that the supergravity fields satisfy the same
equations of motion discovered by \cite{AP}. 
In other words, the same flow equations in 4-dimensions provide
two different 11-dimensional solutions to the equations of the
11-dimensional supergravity.

When
we go to 11-dimensional theory from the 4-dimensional gauged
supergravity, the various
11-dimensional solutions will occur even if the 4-dimensional flow equations are
the same. We expect that since the flow
equations 
in 4-dimensions are related to the ${\cal N}=2$ supersymmetry via $U(1)_R$
symmetry, other types of 
11-dimensional solutions with common 4-dimensional flow equations will
arise. 
One possibility, as we mentioned in \cite{AW0908}, 
is characterized by the $SU(2) \times U(1)
\times U(1)_R$ symmetry which is smaller than $SU(2) \times SU(2)
\times U(1)_R$. The symmetry breaking to $SU(2) \times U(1) \times U(1)_R$ can 
occur from either the $SU(3) \times U(1)_R$ symmetry or $SU(2) \times SU(2)
\times U(1)_R$ symmetry. The metric corresponding to ${\bf CP}^1
\times {\bf CP}^1$ should preserve only 
one of two ${\bf CP}^1$'s symmetries due to the single $SU(2)$ symmetry.

In this paper, we would like to construct 
a new 11-dimensional solution preserving the above $SU(2) \times U(1)
\times U(1)_R$ symmetry. By assuming that the $AdS_4$ supergravity
fields satisfy the supersymmetric RG flow equations,  
we should find out the correct 7-dimensional internal space possessing
this global symmetry. 
By realizing that the five-dimensional Sasaki-Einstein $T^{1,1}$ space
can be generalized to the 5-dimensional Sasaki-Einstein $Y^{p,q}$ 
space \cite{GMSW} where $p$ and $q$ are positive integers with $0\leq
q \leq p$,
it is obvious to consider this space first. 
When $p=1$ and $q=0$, the $Y^{1,0}$ space is nothing but $T^{1,1}$ 
space and moreover the isometry of $Y^{p,q}$ is identical to the above
$SU(2) \times U(1) \times U(1)_R$.
The main procedure given in \cite{CPW} is to start with the round
compactification in terms of $U(1)$-fibration over the Einstein-Kahler
3-fold,
to squash this Einstein-Kahler base ellipsoidally, to stretch the
$U(1)$ fiber, and to introduce 3-form tensor gauge potential
proportional to the volume form on the base. 
Inside of Einstein-Kahler 3-fold, one had either ${\bf CP}^2$ space or ${\bf
CP}^1 \times {\bf CP}^1$ space. Are there any other Einstein-Kahler 2-folds?

Fortunately, in the construction of $Y^{p,q}$ space, it is known that 
$Y^{p,q}$ space can be written in terms of $U(1)$ bundle over
the Einstein-Kahler 2-fold. Therefore, there is a room for this
4-dimensional Einstein-Kahler 2-fold inside of
above Einstein-Kahler 3-fold. Then the next step is to find out the correct
4-form field strengths in this background. Before we use the
11-dimensional Einstein-Maxwell equations directly, it is better to
imitate the 3-forms appeared in previous ${\bf CP}^2$ or ${\bf CP}^1
\times {\bf CP}^1$ cases. Basically the structure of 3-form from the
triple wedge product between the orthonormal frames looks similar to
each other. The overall functional dependence on the $AdS_4$
supergravity fields and the exponential factors corresponding to
unbroken $U(1)$ symmetries can be determined by solving the 11-dimensional
Einstein-Maxwell equations directly.

In section 2, starting with the two parts of $Y^{p,q}$ space metric,
$U(1)$ bundle and the 4-dimensional base space which is
Einstein-Kahler 2-fold, we put them inside of the squashed and stretched  
7-dimensional internal space appropriately.  Then one determines the
full 11-dimensional metric with the correct warp factor. 
Assuming that the two $AdS_4$ supergravity fields satisfy the domain
wall solutions, one computes the Ricci tensor in this background completely.  
For the 4-form field strengths, one makes an ansatz by writing the
three parts, 1) the overall function, 2) the exponential function with
$U(1)$'s and 3) the triple wedge product between the orthonormal frames.
Eventually, the 11-dimensional Einstein-Maxwell equations determine  
all the undetermined quantities.

In section 3, we summarize the results of this paper and make some
future directions.

In the Appendix, we present the detailed expressions for the Ricci
tensor and 4-form field strengths.

\section{An ${\cal N}=2$ supersymmetric 
$SU(2) \times U(1)  \times U(1)_R$-invariant 
flow in an 11-dimensional theory}

When the 11-dimensional supergravity is reduced to 4-dimensional
${\cal N}=8$ gauged supergravity, the 4-dimensional spacetime metric contains
a warp factor which depends
on both 4-dimensional spacetime coordinates and 
7-dimensional internal space coordinates. The internal metric of deformed
seven-sphere can be obtained from the $AdS_4$ supergravity data, the
supergravity fields $(\rho, \chi)$, using
the explicit formula \cite{dWNW}, and the warp factor is also determined.   
We have
\bea
ds_{11}^2 =\Delta(r,\mu)^{-1} \, \left(dr^2 +e^{2 A(r)}
\, \eta_{\mu\nu}\, dx^\mu dx^\nu \right)+ L^2 \,
\sqrt{\Delta(r,\mu)} \, ds_7^2(\rho,\chi),
\label{11dmetric}
\eea
where the 3-dimensional metric is given by 
$\eta_{\mu\nu}=(-,+,+)$, the radial variable $r=x^4$ is the coordinate 
transverse to
the domain wall, the scale factor $A(r)$ behaves linearly in $r$ at UV
and IR regions, $L$ is a radius of round seven-sphere ${\bf S}^7$ and the warp
factor $\Delta(r,\mu)$ also depends on the $\mu$ that is one
of the internal coordinates($\mu=x^5$) as well as the radial coordinate $r$ via
the supergravity fields $(\rho, \chi)$.

Let us assume that the supergravity fields $(\rho, \chi)$
in 4-dimensions satisfy the supersymmetric RG flow equations \cite{AP} 
with $SU(3) \times
U(1)_R$ symmetry in the convention of \cite{CPW}:
\bea
\frac{d \rho}{d r} & = & \frac{1}{8L \, \rho} \, \left[
  (\cosh(2\chi) +1) + \rho^8\, (\cosh(2\chi)-3) \right],
\nonu \\
\frac{d \chi}{d r} & = & \frac{1}{2L \, \rho^2} \,
(\rho^8-3) \,
\sinh(2\chi),
\nonu \\
\frac{d A}{d r} & = & \frac{1}{4L \, \rho^2} \, 
\left[ 3
(\cosh(2\chi)+1) -\rho^8 (\cosh(2\chi)-3)\right].
\label{domain}
\eea
In 4-dimensions, there exist two critical points, ${\cal N}=8$ $SO(8)$ critical
point at which $(\rho, \chi)=(1,0)$ and ${\cal N}=2$
$SU(3) \times U(1)_R$ critical point at which $(\rho,
\chi)=(3^{\frac{1}{8}}, \frac{1}{2} \cosh^{-1} 2)$. 
One can easily check that at these two points, 
$\frac{d \rho}{d r}$
and $\frac{d \chi}{d r}$ vanish due to the right hand sides of 
(\ref{domain}) are equal to zero.
Furthermore, the criticality can be observed from the fact that the
first two right hand sides of (\ref{domain}) can be written as the
derivatives of superpotential $W(\rho, \chi)$ 
with respect to the field $\rho$ and the
field $\chi$ respectively. 
One can read off the superpotential $W(\rho, \chi)$ explicitly by
realizing that the right hand side of third equation in (\ref{domain}) 
is equal to $-\frac{2}{L} W(\rho, \chi)$.
The superpotential has $1$ and $\frac{3^{\frac{3}{4}}}{2}$ at two
critical values respectively.
We will see the 11-dimensional lift of this superpotential, geometric 
superpotential, later when we discuss about the 11-dimensional field equations.

We need to find out the correct 7-dimensional metric which preserves 
$SU(2) \times U(1) \times U(1)_R$ symmetry which maybe obtained from
the symmetry breaking of above bigger symmetry $SU(3) \times U(1)_R$
corresponding to the stretched five-sphere ${\bf S}^5$ 
described by $U(1)$ bundle over the ${\bf CP}^2$ 
or $SU(2) \times SU(2) \times U(1)_R$ symmetry corresponding to the stretched
$T^{1,1}$
space realized by $U(1)$ bundle over ${\bf CP}^1 \times {\bf CP}^1$.
Once we have found this 7-dimensional internal metric with the warp 
factor given in \cite{CPW}, then 
the full 11-dimensional metric can be written as (\ref{11dmetric}).   
Then how one can find this internal metric with the above 
specific symmetry? 
It is not obvious that the $SU(2)$ symmetry among the full $SU(2) \times
U(1) \times U(1)_R$ symmetry is realized from ${\bf CP}^2$ space which
preserves $SU(3)$ symmetry. However, any ${\bf CP}^1$ factor in 
${\bf CP}^1 \times {\bf CP}^1$ space can provide this $SU(2)$ symmetry
because  the ${\bf CP}^1$ preserves the $SU(2)$ symmetry.
So, our strategy is to look at the second solution of \cite{CPW}
closely
rather than the first solution.

At first, let us replace the 4-dimensional ${\bf CP}^1 \times {\bf
CP}^1$ space appearing in the 7-dimensional internal space in \cite{CPW}
with the 4-dimensional 
Einstein-Kahler 2-fold which lives in the five-dimensional $Y^{p,q}$
space \cite{GMSW}. 
Next, we need to find out the correct one-form which contains 
the $U(1)$ bundle over this Einstein-Kahler 2-fold.
This one-form $\omega$ is given by \footnote{The 11-th coordinate $\alpha$
  here corresponds to $\psi$ introduced in \cite{CPW}.}
\bea
\omega = \frac{1}{2} \, \sin (2\mu) \left[-\frac{1}{\rho(r)^4} \, d \alpha +
  \rho(r)^4 \, (u, J d u)\right],
\label{omega}
\eea
where we introduce the ${\bf R}^8$ vector 
$u=(u^1, \cdots, u^6, 0, 0)$ which parametrize a unit
${\bf S}^5$ sphere and $J$ is the Kahler form
with $J_{12}=J_{34}=J_{56}=J_{78}=1$. The product $(u, J du)$ is
defined as $(u, J d u) \equiv u^A J_{AB} u^B$.
Note that the one-form in subsection 4.1 of 
\cite{CPW} is the $U(1)$ bundle over the
4-dimensional ${\bf CP}^2$ space while the one-form in Eq. (4.38) of 
\cite{CPW} is the $U(1)$ bundle over the
4-dimensional ${\bf CP}^1 \times {\bf CP}^1$ space.

What is $(u, J d u)$ in (\ref{omega}) corresponding to the $U(1)$
bundle over the Einstein-Kahler 2-fold? 
Let us recall the metric for 
the 5-dimensional Sasaki-Einstein space $Y^{p,q}$ with $c=1$ \cite{GMSW}
\bea
ds^2_{Y^{p,q}}  & = & 
ds_{EK(2)}^2 +  \frac{1}{9}\, \left[ (d \psi -\cos \theta \, d \phi) +
  y \, ( d \beta + 
\cos \theta \, d \phi) \right]^2
\nonu \\
&=& 
\left[ \frac{1}{6}\,(1-y) \, (d \theta^2 + \sin^2\theta \,
  d \phi^2) + \frac{1}{w(y) q(y)} \, d y^2 + \frac{1}{36} w(y)
  q(y)\, ( d \beta +\cos \theta \, d \phi)^2 \right] 
\nonu \\
&+& 
 \frac{1}{9}\, \left[ (d \psi -\cos \theta \, d \phi) +  y \, ( d \beta + 
\cos \theta \, d \phi) \right]^2,
\label{ypq}
\eea
where $y$-dependent functions are given by
\bea
w(y) \equiv \frac{2(a-y^2)}{1-y}, \qquad q(y) 
\equiv \frac{a-3y^2 + 2 y^3}{a-y^2}, \qquad
a = \frac{1}{2} -\frac{(p^2-3q^2)}{4p^3} \sqrt{4p^2 -3 q^2}.
\label{wqa}
\eea
Also note that the form in the last line of (\ref{ypq}) provides the
Kahler 2-form and satisfies
\bea
\frac{1}{6} d  \left[ -\cos \theta \, d \phi +  y ( d \beta + 
\cos \theta \, d \phi) \right] = \frac{1}{6}\, (1-y)\, \sin \theta \,d
\theta \wedge d \phi + \frac{1}{6} \, d y \wedge 
( d \beta + 
\cos \theta \, d \phi). 
\label{Kahler2}
\eea
Then it is natural to view that 
we identify  $(u, J d u)$ with the $U(1)$ bundle over 
this Einstein-Kahler 2-fold as follows:
\bea
(u, J d u) =
 \frac{1}{3}\left[ (d \psi -\cos \theta \, d \phi) +  y \, ( d \beta + 
\cos \theta \, d \phi) \right].
\label{uJdu}
\eea 
By plugging (\ref{uJdu}) into (\ref{omega}), we have one-form $\omega$ explicitly.

Finally, we should write down the $U(1)$ Hopf fiber $(x, J d x)$
on ${\bf CP}^3$(note that for $\rho=1$ and $\chi=0$, the internal
metric should contain a ${\bf CP}^3$ factor) where
$x=(x^1,\cdots,x^8)$ is a vector on ${\bf R}^8$
in terms of
$(u, J d u)$ using \cite{CPW}
\bea
(x, J d x) = \cos^2\mu \, (u, J d u) + \sin^2\mu \, d \alpha.
\label{xJdx}
\eea
One also introduces another vector $v=(0,\cdots, 0, \cos \alpha, \sin
\alpha)$ in ${\bf R}^8$ and then the above $x$ can be written as 
$x=u \cos \mu + v \sin \mu$. In (\ref{xJdx}), we used 
$(v, J d v) = d\alpha$.
The 7-dimensional internal space metric $ds^2(\rho, \chi)$ without 
a warp factor can be 
written as
\bea
&& ds_7^2(\rho, \chi)   =  \rho(r)^{-4}  \xi^2  d\mu^2 + \rho(r)^2  \cos^2\mu
ds^2_{EK(2)} + \xi^{-2} \omega^2 +\xi^{-2} \cosh^2\chi(r)  (x, J
d x)^2.
\label{7d}
\eea
Here we substituted  the metric (\ref{ypq}) for the Einstein-Kahler 2-fold where
the four coordinates are parametrized by 
$(\theta =x^6,\phi=x^7,y=x^8,\beta=x^9)$ in the
second term of (\ref{7d}).
After plugging the 1-form (\ref{omega}) with (\ref{uJdu}) in the third
term of (\ref{7d}) and the
$U(1)$ Hopf fiber (\ref{xJdx}) in the last term of (\ref{7d}), 
we obtain the final 7-dimensional
internal metric preserving $SU(2) \times U(1) \times U(1)_R$ symmetry
as follows:
\bea
&& ds_7^2(\rho, \chi) =    \rho(r)^{-4} \, \xi(r,\mu)^2 \, d\mu^2 \nonu \\
&& +  
\rho(r)^2 \, \cos^2\mu
\, \left[ \frac{1}{6}\,(1-y) \, (d \theta^2 + \sin^2\theta \,
  d \phi^2) + \frac{1}{w(y) q(y)} \, d y^2 + \frac{1}{36} w(y)
  q(y)\, ( d \beta +\cos \theta d \phi)^2 \right] 
\nonu \\
&&+  \xi(r,\mu)^{-2} \,  \frac{1}{4} \, \sin^2 (2\mu) \,
\left[-\frac{1}{\rho(r)^4} \, d \alpha +
  \rho(r)^4 \,  \frac{1}{3}\, \left[ (d \psi -\cos \theta \, d \phi) +  y
    \, ( d \beta + 
\cos \theta \, d \phi) \right] \right]^2
\nonu \\
&& +  \xi(r,\mu)^{-2} \, \cosh^2\chi(r) \, \left[ \sin^2\mu \, d \alpha +
\cos^2\mu \,  \frac{1}{3}\, \left[ (d \psi -\cos \theta \, d \phi) +  
y \, ( d \beta + 
\cos \theta \, d \phi) \right] \right]^2,
\label{7dmetric}
\eea
where the quadratic form $\xi^2 \equiv (x, Q x)$ with
$Q=\mbox{diag}(\rho(r)^{-2},\cdots,
\rho(r)^{-2}, \rho(r)^{6}, \rho(r)^{6})$ in 8-dimensional space 
can be computed and it is given by \cite{CPW}
\bea
\xi(r,\mu) = \frac{\sqrt{X(r,\mu)}}{\rho(r)}, \qquad
X(r,\mu) \equiv \cos^2\mu + \rho(r)^8 \, \sin^2\mu.
\nonu
\eea 
In (\ref{7dmetric}), we explicitly presented the $r$-dependence in
every place. 
The nontrivial squashing characterized by $\rho(r)$ deforms the metric 
on the ${\bf CP}^3$(by changing the variables appropriately \cite{GMSW} one makes 
the 5-dimensional metric on $Y^{p,q}$ space as a $U(1)$ bundle over 
the Fubini-Study  
metric on ${\bf CP}^2$, one obtains the usual round 5-sphere ${\bf
  S}^5$ and the first three lines of (\ref{7dmetric}) contain ${\bf CP}^3$ metric) 
and moreover rescales the Hopf fiber which appears in the last line of
(\ref{7dmetric}). The stretching is characterized by $\chi(r)$.
However, there exists $SU(2) \times U(1)$ symmetry
from the structure of Einstein-Kahler 2-fold in $ds^2_{EK(2)}$.
The $U(1)$ symmetry is generated by the angle $\beta$.
The combined two $U(1)$ symmetries by the angle $\psi(=x^{10})$ and the angle
$\alpha (=x^{11})$
will provide a single $U(1)_R$ symmetry which is relevant to the
${\cal N}=2$ supersymmetry. We will return to this issue when we
discuss about the 4-form field strengths later.

For $\mu=0$, the 7-dimensional metric (\ref{7dmetric}) 
reduces to the following metric on moduli space for the M2-brane probe
\bea
\rho(r)^2 ds^2_{Y^{p,q}} +\rho(r)^2 \sinh^2 \chi(r) \,
\frac{1}{9} \left[ (d \psi -\cos \theta \, d \phi) +  y ( d \beta + 
\cos \theta \, d \phi) \right]^2,
\label{modu}
\eea
where the metric for ${Y^{p,q}}$ is given by (\ref{ypq}). 
In particular, the ${\bf S}^5$ or $T^{1,1}$ is replaced by $Y^{p,q}$
and for large $r$ the moduli space (\ref{modu}) approaches the Ricci-flat conifold. 
Now one sees that the function  $\sinh^2 \chi(r)$ plays the role of a
stretching of the $U(1)$-fiber. Then on can say, for this particular
coordinate $\mu=0$, there exists 
a stretched five-sphere ${\bf S}^5$, a stretched
$T^{1,1}$ space or a stretched $Y^{p,q}$ space depending on the $U(1)$-fibers.

From these observations so far, 
we obtain the following set of frames for the
11-dimensional metric (\ref{11dmetric}):
\bea
e^1  & = & -\frac{1}{\sqrt{\Delta(r,\mu)}} \, e^{A(r)} \, d x^1,
\qquad
e^2   =  \frac{1}{\sqrt{\Delta(r,\mu)}} \, e^{A(r)} \, d x^2,
\qquad
e^3   =  \frac{1}{\sqrt{\Delta(r,\mu)}} \, e^{A(r)} \, d x^3,
\nonu \\
e^4   &  = &  \frac{1}{\sqrt{\Delta(r,\mu)}}  \, d r,
\nonu \\
e^5   & = &  L \,
\Delta(r,\mu)^{\frac{1}{4}} \, \frac{\sqrt{X(r,\mu)}}{\rho(r)^3} \, d \mu,
\nonu \\
 e^6  &  = &  L \,
\Delta(r,\mu)^{\frac{1}{4}} \,\rho(r) \cos \mu \, \sqrt{\frac{1-y}{6}} \, d \theta,
\nonu \\
 e^7 & = & L \,
\Delta(r,\mu)^{\frac{1}{4}} \,\rho(r) \cos \mu \, \sqrt{\frac{1-y}{6}} \, \sin \theta
\, d \phi,
\nonu \\
 e^8  & = &  L \,
\Delta(r,\mu)^{\frac{1}{4}} \,\rho(r) \cos \mu \, \frac{1}{\sqrt{w(y)\,q(y)}} \, d y,
\nonu \\
 e^9 & = & L \,
\Delta(r,\mu)^{\frac{1}{4}} \,\rho(r) \cos \mu \, \frac{1}{6} \, \sqrt{w(y)\,q(y)} \, 
(d\, \beta + \cos \theta \, d \phi),
\label{11frames}
\\
 e^{10} & = & L \,
\Delta(r,\mu)^{\frac{1}{4}} \, \frac{\rho(r)}{\sqrt{X(r,\mu)}} \, \frac{1}{2} \, \sin(2\mu) \, 
\left[ - \frac{d \alpha}{\rho(r)^4}
  + \frac{1}{3}\, \rho(r)^4\, \left[ (d \psi -\cos \theta \, d \phi) +  y
    \, ( d \beta + 
\cos \theta \, d \phi) \right] \right],
\nonu
\\
 e^{11} & = & L \,
\Delta(r,\mu)^{\frac{1}{4}} \,  \frac{\rho(r) \, \cosh \chi(r)}{\sqrt{X(r,\mu)}} \, 
\left[ \sin^2\mu \, d \alpha + 
\frac{1}{3} \cos^2 \mu \, \left[ (d \psi -\cos \theta \, d \phi) + y\, ( d \beta + 
\cos \theta \, d \phi) \right] \right],
\nonu
\eea
where the warp factor is given by \cite{CPW}
\bea
\Delta(r,\mu) = \frac{\rho(r)^{\frac{4}{3}}}{ X(r,\mu)^{\frac{2}{3}}
  \, 
\cosh^{\frac{4}{3}} \chi(r) }.
\label{delta}
\eea
The constant $L$ in (\ref{11frames}) in the 7-dimensional internal space 
is determined by using the symmetry of UV fixed point later. 

Denoting the 11-dimensional metric as $g_{MN}$ with 
the convention 
$(-, +, \cdots, +)$
and the antisymmetric 
tensor fields as $F_{MNPQ}$, the
Einstein-Maxwell equations are given by \cite{CJS}
\bea
R_{M}^{\;\;\;N} & = & \frac{1}{3} \,F_{MPQR} F^{NPQR}
-\frac{1}{36} \de^{N}_{M} \,F_{PQRS} F^{PQRS},
\nonu \\
\nabla_M F^{MNPQ} & = & -\frac{1}{576} \,E \,\ep^{NPQRSTUVWXY}
F_{RSTU} F_{VWXY},
\label{fieldequations}
\eea
where the covariant derivative $\nabla_M$ 
on $F^{MNPQ}$ in 
(\ref{fieldequations})
is given by 
$E^{-1} \pa_M ( E F^{MNPQ} )$ together with elfbein determinant 
$E \equiv \sqrt{-g_{11}}$. The epsilon tensor 
 $\ep_{NPQRSTUVWXY}$ with lower indices is purely numerical.
All the indices are based on the coordinate basis.

$\bullet$ At  the $SO(8)$-invariant UV fixed point \cite{AP}

For 
\bea
\rho(r) = 1, \qquad \chi(r) =0,
\label{rhochi}
\eea
one should recover the maximally symmetric $AdS_4 \times {\bf S}^7$ solution.
In general, one can introduce the arbitrary coefficients 
in the frames $e^6$ to $e^{11}$ of (\ref{11frames}). But 
these can be fixed in order to make the Ricci tensor have the form 
\bea
R_M^{\,\,N} =\frac{6}{L^2} \diag (-2, -2, -2, -2,
1, 1, 1, 1, 1, 1, 1 ),
\nonu
\eea 
which fixes the round ${\bf S}^7$ radius to be $L$, twice the $AdS_4$
radius, as expected. 
As Freund-Rubin parametrization \cite{FR}, the 3-form gauge field with 
3-dimensional M2-brane indices maybe defined by \cite{CPW}
\bea
A^{(3)} = \frac{1}{2} e^{\frac{6r}{L}} \, d x^1
\wedge d x^2 \wedge d x^3.
\label{A3}
\eea
Note that at the UV end of the flow the function $A(r)$ behaves as
$\frac{2}{L} r$ from the solution (\ref{domain}) for $A(r)$ and $W=1$. 
The exponential factor $e^{3A(r)}$ will be compensated by the same
factor from the 11-dimensional metric when we derive the geometric
superpotential along the flow.
From (\ref{A3}), one obtains the only nonzero component for the 4-form 
as $ F_{1234} =-\frac{18}{L}$.

$\bullet$ At  the $SU(2) \times U(1) \times U(1)_R$-invariant 
IR fixed point \cite{AP}

As we mentioned before, there exists IR critical point characterized 
by 
\bea
\rho(r) = 3^{\frac{1}{8}}, \qquad \chi(r) =\frac{1}{2} \cosh^{-1} 2.
\label{rhochi1}
\eea
The 3-form gauge field with 
3-dimensional M2-brane indices can be constructed as the UV critical
point.
By realizing that  at the IR end of the flow the function $A(r)$ behaves as
$\frac{3^{\frac{3}{4}}}{L} r$ from the solution (\ref{domain}) for
$A(r)$ and $W=
\frac{3^{\frac{3}{4}}}{2}$(and we define $\hat{L} \equiv
3^{-\frac{3}{4}} L$), one writes down the 3-form gauge field including
the internal parts as follows \cite{CPW}:
\bea
A^{(3)} = \frac{3^{\frac{3}{4}}}{4} e^{\frac{3r}{\hat{L}}} \, d x^1
\wedge d x^2 \wedge d x^3 + C^{(3)} + (C^{(3)})^{\ast}.
\label{a3}
\eea
How does one determine the internal 3-form field $C^{(3)}$?
The Kahler form in (\ref{Kahler2}) contains 
$e^6 \wedge e^7$ and $e^8 \wedge e^9$ that lead to
the natural basis of the one-forms and the ${\bf CP}^3$ factor for
$\rho=1$ and $\chi=0$ has also $e^5$ and $e^{10}$ which can be
combined together.
In fact, we find  
\bea
C^{(3)} = -\frac{1}{4} \sinh \chi(r) \, e^{i(\alpha+\psi)} \, (e^5- i e^{10}) 
\wedge (e^6 + i e^7) \wedge (e^8 + i e^9).
\label{c3}
\eea
In general, the overall function depends on both $\rho(r)$ and $\chi(r)$.
However, the above expression (\ref{c3}) possesses only $\chi(r)$-dependence.
The coefficients for $\alpha$ and $\psi$ in the exponent are fixed as
$1$ and $1$ respectively. 
We have considered the angle $\beta$ also in the exponent but the
coefficient for this vanishes from 11-dimensional Einstein equation.
Although the structure of triple product (\ref{c3}) 
between the orthonormal basis looks very similar to the previous 
constructions with $SU(3) \times U(1)_R$ symmetry or $SU(2) \times SU(2) \times
U(1)_R$ symmetry(up to signs), the functional behavior of the
exponential function, i.e., the rotations with $U(1)$ symmetries 
in the fields behave differently.   
It is interesting note that the overall function contains
$\sinh \chi(r)$ which plays the role of a stretching $U(1)$ fiber we
described before.

Let us explain all these in detail. Let us go to the Ricci tensor
first in the frame basis we introduced in (\ref{11frames}).
The Ricci tensor has only two nonvanishing off-diagonal
components:$R_{10}^{\,11}$ and $R_{11}^{\,10}$. There exists a
nontrivial identity between these components. 
It turns out the Ricci tensor is identical to the one with $SU(3)
\times U(1)_R$ symmetry 
or the one with $SU(2) \times SU(2) \times U(1)_R$ symmetry.
That is, the Ricci tensor for three cases 
has same value(in the frame basis) at the IR critical point.
Let us present them here for convenience:
\bea
R_1^{\, 1} & = & -\frac{(55-32 \cos 2\mu + 3 \cos 4\mu) }
{3 \cdot 2^{\frac{1}{3}} \,\sqrt{3} \, \hat{L}^2\,
  (2-\cos 2\mu )^{\frac{8}{3}}} =R_2^{\, 2} =R_3^{\, 3}=R_4^{\, 4}
= -2 R_6^{\, 6} = -2 R_7^{\, 7} = -2
R_8^{\, 8}=-2 R_9^{\, 9},
\nonu \\
R_5^{\, 5} & = & \frac{(29-16 \cos 2\mu) }
{3 \cdot 2^{\frac{1}{3}} \,\sqrt{3} \,\hat{L}^2\,
  (2-\cos 2\mu )^{\frac{8}{3}}} = R_{10}^{\, 10}, \qquad
R_{10}^{\, 11} =  -\frac{2 \cdot 2^{\frac{1}{6}} \sin 2\mu }
{\sqrt{3} \,\hat{L}^2\,
  (2-\cos 2\mu )^{\frac{5}{3}}} = R_{11}^{\,\,10},
\nonu \\
R_{11}^{\, 11} & = & \frac{(80-64 \cos 2\mu +9 \cos 4 \mu) }
{3 \cdot 2^{\frac{1}{3}} \,\sqrt{3} \,\hat{L}^2\,
  (2-\cos 2\mu )^{\frac{8}{3}}}.
\label{Ricci}
\eea
All these depend on only $\mu(=x^5)$ coordinate. One can also obtain the
Ricci tensor in coordinate basis that depends on $y(=x^8)$ 
and $\theta(=x^6)$ as well as $\mu$. 
Now it is ready to use the 11-dimensional Einstein equation which is
the first one of (\ref{fieldequations}) where the indices are based on
the coordinate basis. One can transform this Einstein equation with
coordinate basis into the one with frame basis without any difficulty
via (\ref{11frames}). 
The $(10, 9)$ component of 
right hand side of Einstein equation is nonzero in general but 
the corresponding $R_{10}^{\,9}$ from (\ref{Ricci}), which
appears in the left hand side of Einstein equation, vanishes. This implies 
that the coefficient of $\beta$ should vanish and the coefficient of
$\psi$ should be $1$ in the exponent of 3-form (\ref{c3}).
Then there exists a $U(1)$ symmetry generated by the angle $\beta$.
Furthermore, by comparing the $(10, 11)$  component of Einstein
equation,
the coefficient for the angle $\alpha$ which is equal to $1$ 
and the overall coefficient of
3-form that is $-\frac{1}{4}$ are completely fixed. 
At the moment, one cannot determine the functional dependence for 
$\sinh\chi(r)$ in (\ref{c3}) because we are looking for the behavior
at the critical point (\ref{rhochi1}). We return to this issue when we
discuss about the RG flow later. 

The internal part of $F^{(4)}$ can be written as
$ d  C^{(3)} + d (C^{(3)})^{\ast}$.
The antisymmetric 
tensor fields can be obtained  from 
$F^{(4)} = d A^{(3)}$ with (\ref{a3}).
It turns out that the antisymmetric 
field strengths have the following nonzero components
in the orthonormal frame basis used in (\ref{c3}) or in
(\ref{11frames})
\bea
F_{1234}  & = & -\frac{3 \cdot  2^{\frac{1}{3}} \cdot 3^{\frac{3}{4}}}
{\hat{L}(2-\cos2\mu)^{\frac{4}{3}}}, \qquad
F_{568\,10} = \frac{2^{\frac{1}{3}} \cdot 3^{\frac{3}{4}}
  \sin(\alpha+\psi) \sin2\mu}
{\hat{L}(2-\cos2\mu)^{\frac{4}{3}}},
=-F_{579\,10}, 
\nonu \\
F_{568\,11} & = & -\frac{2^{\frac{5}{6}} \cdot 3^{\frac{3}{4}} \sin(\alpha+\psi)}
{\hat{L}(2-\cos
  2\mu)^{\frac{1}{3}}}=-F_{579\,11}=F_{69\,10\,11}=F_{78\,10\,11}, 
\nonu \\
F_{569\,10}  & = &  \frac{2^{\frac{1}{3}} \cdot 3^{\frac{3}{4}} 
\cos(\alpha+\psi) \sin2\mu}
{\hat{L}(2-\cos 2\mu)^{\frac{4}{3}}}=F_{578\,10},
\nonu \\
F_{569\,11} & = & -\frac{2^{\frac{5}{6}} \cdot 3^{\frac{3}{4}} \cos(\alpha+\psi)}
{\hat{L}(2-\cos
  2\mu)^{\frac{1}{3}}}=F_{578\,11}=-F_{68\,10\,11}=F_{79\,10\,11}, 
\label{F4}
\eea
where the angle-dependences for $\alpha$ and $\psi$ appear in the
combination of $(\alpha + \psi)$ as observed previously. 
One can make the two $U(1)$ symmetries 
generated by $\alpha$ and $\psi$ 
which preserve this combination $(\alpha +\psi)$.
Note that these 4-forms break the $SU(2) \times U(1)_{\beta} \times U(1)_{\alpha}
\times U(1)_{\psi}$ into $SU(2) \times U(1) \times U(1)_R$ where $U(1)$ is
generated by the angle $\beta$. It is obvious that the invariance of 
$U(1)_{\beta}$ comes from the fact that (\ref{F4}) do not depend on
the angle $\beta$ as we explained before.
After substituting (\ref{F4}) into 
the right hand side of Einstein equation (\ref{fieldequations})
with frame basis (\ref{11frames}) 
one reproduces the one of $SU(3) \times U(1)_R$ case
\cite{CPW} or $SU(2) \times SU(2) \times U(1)_R$ exactly. 
This feature is also expected since as we already mentioned, the Ricci
tensor for three independent cases is identical to each other.
In other words, the 4-forms themselves are different from each other, 
their combinations appearing in the right hand side of Einstein equation
are the same.
In particular, the 4-form given in (\ref{F4}) looks very similar to 
the one of $SU(2) \times SU(2) \times U(1)_R$ symmetry case: same
independent components up to signs. 

$\bullet$ Along the  the $SU(2) \times U(1) \times U(1)_R$-invariant
RG flow

The nontrivial $r$-dependence of supergravity fields $(\rho,\chi)$ via 
(\ref{domain}) requires that the 11-dimensional Einstein-Maxwell 
  equations become consistent with not only at the critical points but
  also along the supersymmetric RG flow connecting the two critical
  points.
For solutions with varying scalars, the ansatz for the 4-form field
strength will be more complicated. 
We will apply the correct ansatz for the 11-dimensional 3-form gauge
field by acquiring the $r$-dependence of the supergravity scalars and
will derive the 11-dimensional Einstein-Maxwell equations
corresponding to the $SU(2) \times U(1) \times U(1)_R$-invariant RG
flow.

Let us take the 3-form ansatz as follows \cite{CPW}: 
\bea
A^{(3)} = \widetilde{W}(r,\mu) \, e^{3A(r)} \, d x^1
\wedge d x^2 \wedge d x^3 + C^{(3)} + (C^{(3)})^{\ast},
\label{a3flow}
\eea
where $ C^{(3)}$ is given by (\ref{c3}) as before.
Then how does one determine the $\chi(r)$ dependence appearing this 3-form? 
One puts an arbitrary function $f(\rho(r),\chi(r))$ in front of this 3-form
at the beginning.
One also obtains the Ricci tensor from the 11-dimensional 
metric (\ref{11dmetric})
when the supergravity fields $(\rho(r), \chi(r))$ vary with respect to the
$r$-coordinate. They are given in (\ref{Ricci1}) of 
the Appendix A where all the derivative terms before using the flow
equations disappear by constraining the conditions (\ref{domain}).
When one needs to have the second derivative terms for 
$\rho(r), \chi(r)$ or $A(r)$, one should differentiate the flow equations
further and change the right hand side by using the flow equations
again and removing the derivative terms.  
The $(10, 11)$ component of Einstein equation determines the function
$f(\rho(r), \chi(r))$. The $R_{10}^{\, 11}$ component is given in
(\ref{Ricci1}) while the corresponding right hand side depends on this
function and its derivative. One obtains $v(r) f(v(r)) +(1-v(r)^2)f'(v(r))=0$
where $v(r) \equiv \cosh\chi(r)$. This implies that the solution
$f(v(r))$ is exactly the same as $\sinh\chi(r)$.

Now let us determine the exact form for the geometric superpotential
introduced in (\ref{a3flow}).
Let us consider $(4, 4), (4, 5)$ and $(5, 5)$ components of Einstein equation.
The first and last ones contain 
$
\widetilde{W}^2, 
\widetilde{W}  \pa_{r}  \widetilde{W}, 
(\pa_{r}  \widetilde{W})^2$
and $(\pa_{\mu}  \widetilde{W})^2$
while the second one contains
$
\widetilde{W}   \pa_{\mu}  \widetilde{W}$ 
and
$
\pa_{r}  \widetilde{W} 
 \pa_{\mu}  \widetilde{W}$.
By eliminating $(\pa_{r}  \widetilde{W})^2 $  
from $(4,4)$ and $(5,5)$
components,
one obtains
\bea
\frac{\widetilde{W}(r,\mu)}{\pa \mu} = 
-\frac{1}{2\rho(r)^2} \left[
\cosh^2 \chi(r) + \rho(r)^8(-2 + \cosh^2 \chi(r))\right] \sin2\mu.
\label{muw}
\eea
By integrating this (\ref{muw}) with respect to the $\mu$ coordinate, one gets
\bea
\widetilde{W}(r,\mu) = \frac{1}{4\rho(r)^2} \left[
\cosh^2 \chi(r) + \rho(r)^8(-2 + \cosh^2 \chi(r))\right] \cos2\mu + g(r),
\label{w}
\eea 
where $g(r)$ is an arbitrary function of $r$. How one can determine
the function $g(r)$? By recalling that the
superpotential $W(\rho,\chi)$ in 4-dimensions has terms like
$\rho(r)^6$ or $\rho(r)^{-2}$. Then one makes further ansatz for
$g(r)$ as $g(r) = \rho(r)^{-2} \, h_1(\chi(r)) + \rho(r)^6 \, h_2(\chi(r))$. 
Let us insert these into the $(4,5)$ component of Einstein equation.
Then the unknown functions $h_1(\chi(r))$ and $h_2(\chi(r))$ are
completely fixed and they are given by 
\bea
h_1(\chi(r)) = \frac{1}{4} \cosh^2 \chi(r), \qquad 
h_2(\chi(r)) = \frac{1}{8} \left(3-\cosh2\chi(r) \right).
\label{h}
\eea
By plugging these (\ref{h}) into (\ref{w}) with $g(r)$ above, 
one obtains the final expression for the geometric superpotential
as follows:
\bea
 \widetilde{W}(r,\mu) = \frac{1}{4\rho(r)^2} \left[ 
\left(\cosh2\chi(r) +1 \right) \cos^2 \mu - \rho(r)^8 \left( \cosh2\chi(r)-
 3 \right) \sin^2 \mu \right],
\label{geometric}
\eea
which is exactly the same as the one \cite{CPW} found in other two cases.
When $\cos^2 \mu =\frac{3}{4}$, then we have $ \widetilde{W}(r,\mu)
=-\frac{1}{2} W(\rho, \chi)$ where $W(\rho,\chi)$ is a superpotential
in 4-dimensions.

Comparing with the previous 4-form fields at the IR fixed point, 
the mixed 4-form fields $F_{\mu\nu\rho 5}, F_{4mnp}$ and $F_{45mn}$
where $\mu, \nu, \rho =1, 2, 3$ and $m,n, p =6, 7, \cdots, 11$ are
new if we look at the (\ref{F41}). Indeed, they are not forbidden 
to occur by the global symmetry once we suppose that the 4-dimensional
metric has the domain wall factor $e^{3A(r)}$ that breaks the
4-dimensional conformal invariance.  At both UV and IR critical points,
the 4-dimensional spacetime becomes asymptotically $AdS_4$ which has
conformal invariance and the
mixed field strengths should vanish there. 

In order to check the remaining Maxwell equation, one needs to know
the elfbein determinant $E=\sqrt{-g_{11}}$ and it is given by
\bea
E = \frac{1}{4 \, \rho(r)^{\frac{4}{3}}} 9 \cdot \, 3^{\frac{1}{4}} \, e^{3A(r)}
\, \hat{L}^7 \, \cosh^{\frac{4}{3}} \chi(r)  \, (y-1) \, \cos^5\mu \,
\, \sin \theta \, \sin
\mu \,
\left( \cos^2\mu + \rho(r)^8 \, \sin^2\mu\right)^{\frac{2}{3}},
\nonu
\eea
by computing the determinant of 11-dimensional metric (\ref{11dmetric}).
The right hand side of Maxwell equation of (\ref{fieldequations})
contains also the determinant of 11-dimensional inverse metric. 
Written in terms of coordinate basis, one should also transform the
4-forms in (\ref{F41}) with frame basis of the Appendix B into the
ones with coordinate basis via $e^a_m$ appearing in (\ref{11frames}).
On the other hand, the left hand side of Maxwell equation has 4-form
with upper indices which can be determined by using the 11-dimensional 
inverse metric and 4-forms with lower indices in the coordinate basis.  
We do not present them here because they are rather complicated.
We have checked that all of the Maxwell equations of 
motion are indeed satisfied.

Thus we have established that the solutions (\ref{a3flow}), (\ref{c3}), and 
(\ref{geometric}) actually consists of an exact solution to the
11-dimensional supergravity characterized by bosonic field equations
(\ref{fieldequations}), 
provided that the deformation parameters
$(\rho(r), \chi(r))$ of the 7-dimensional internal space and the
domain wall amplitude $A(r)$ develop in the $AdS_4$ radial direction   
along the $SU(2) \times U(1) \times U(1)_R$-invariant RG flow (\ref{domain}).

So far, we have focused on $c=1$(in other words, for $c\neq 0$ one can
rescale $y$ to set $c=1$ and the metric has one parameter family
characterized by $a$) in the ${\bf S}^2$ metric of
(\ref{ypq})
and the coefficient is given by $(1-y)$. For $c=0$ where $a$ is a
trivial rescaling parameter, then the metric of 
(\ref{ypq}) leads to the standard metric of $T^{1,1}$ space. Then one
can follow the procedure for the second solution in \cite{CPW}.
On the other hand, for $a=1$ where $a$ is defined in (\ref{wqa}) and
$c$ is a trivial rescaling parameter, 
the metric provides the round five-sphere ${\bf S}^5$ metric.
Then one takes the first solution of \cite{CPW}.
Schematically, we draw these solutions in Figure 1.
In 11-dimensional view point, the three independent RG flows
characterized by 
\bea
{\bf S}^5-\mbox{flow} & : & SU(3) \times U(1)_R,  \nonu \\
T^{1,1}-\mbox{flow} & : & SU(2) \times SU(2) \times U(1)_R, \nonu \\
Y^{p,q}-\mbox{flow}& : & SU(2) \times U(1) \times U(1)_R,
\nonu
\eea
arrive at the IR
fixed point at which they have common Ricci tensor (\ref{Ricci}).
Depending on their global symmetry, the internal 3-forms, in each case,  
have the right structures in the exponential function with common 
$\sinh\chi(r)$ dependence. However, the 3-form in the M2-brane 
world-volume directions with the same geometric superpotential 
(\ref{geometric}) 
is common to three different solutions. 
It is surprising that although the 4-forms are different from 
each other completely, the squares of these 4-forms appearing in the
right hand  side of Einstein equation (\ref{fieldequations}) give
rise to the same expressions.  

\begin{figure}[ht]
   \epsfxsize=3.5in 
\centerline{\epsffile{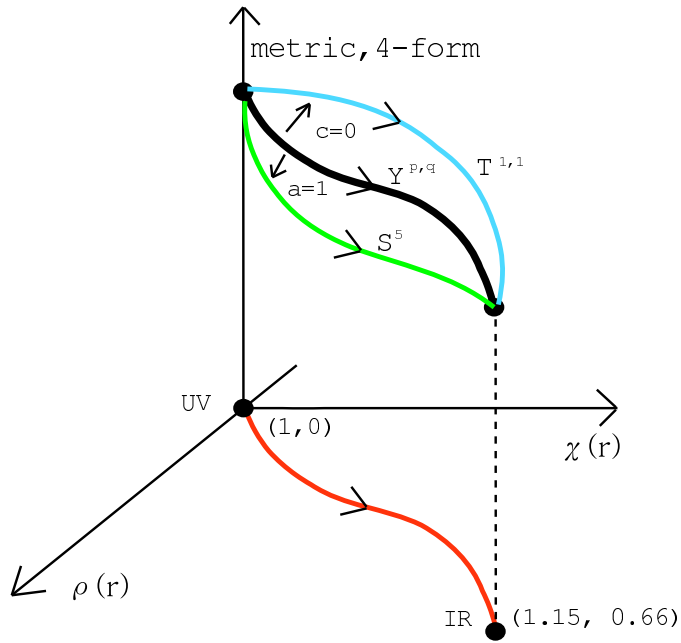}}
   \caption[FIG. \arabic{figure}.]{ 
\sl The RG flow starting from $SO(8)$ UV fixed point to
$SU(3) \times U(1)_R$ IR fixed point and its three 
11-dimensional
lifts.
The theory(in lower plane) 
flows a $SU(3) \times U(1)_R$-invariant fixed point at
which $(\rho,\chi)=(1.15,0.66)$  as
they vary with respect to $r$ according to
(\ref{domain}) \cite{AP}
starting from $(\rho,\chi)=(1,0)$. 
In its 11-dimensional lift, there exist three flows.
At each curve, the 11-dimensional metric and 4-forms vary with the
deformation
parameters $(\rho,\chi)$. 
The solutions to the 
upper and lower ones 
were found in \cite{CPW} while the 
solution to the middle one is found in this paper. The Ricci tensor for
three curves is the same and given in the Appendix A. 
The common $U(1)_R$ factor in the global
symmetry plays the role of ${\cal N}=2$ supersymmetry along the whole
three flows.  Either ${\bf S}^5$-flow or $T^{1,1}$-flow can be
obtained from the more general $Y^{p,q}$-flow by taking the limit
$a=1$ or $c=0$ respectively. }
\end{figure}

\section{
Conclusions and outlook }

We have derived the 11-dimensional Einstein-Maxwell equations
corresponding to the ${\cal N}=2$ $SU(2) \times U(1) \times
U(1)_R$-invariant
RG flow in the 4-dimensional gauged supergravity. The $AdS_4$
supergravity fields can be interpreted as the geometric parameters for
the 7-dimensional internal space. Provided that the $r$-dependence of
these fields is controlled by the RG flow equations, we have found 
the exact solution to the 11-dimensional field equations. With this
solution,
one would say that the $SU(2) \times U(1) \times U(1)_R$-invariant
holographic RG flow can be lifted to an ${\cal N}=2$ M2-brane flow in
M-theory.
The field strengths must be subject to the nontrivial boundary
conditions at both UV and IR critical points.   

It is natural to ask what is corresponding dual gauge theory for the
previous 11-dimensional background in the context of AdS/CFT.
According to the observation of $T^{1,1}$-flow \cite{AW0908},  
the quiver $U(N)^3$ Chern-Simons gauge theory for the M2-branes
probing the cone over $Q^{1,1,1}$ space provides the quiver diagram
for a partial resolution \cite{FKR} of $Q^{1,1,1}$ theory with $U(N)^3$ gauge
group and two $SU(2)$ doublets and an adjoint field.  
It is known that in \cite{GMSW1}, the higher dimensional analog of the 
$Y^{p,q}$ space was found(and denoted by $X^{p,q}$) and can be
expressed as a $U(1)$ bundle over 6-dimensional Einstein-Kahler 
space which is a 2-bundle over a 4-dimensional Einstein-Kahler 
space. Therefore, the partial resolution of the $X^{p,q}$ might be a
candidate for the dual gauge theory. 
The spin-2 KK modes around a warped product of $AdS_4$ and a squashed
and stretched 7-manifold can be obtained. The mass-squared in 
$AdS_4$, in principle, can be determined  
and it is an open problem to find out what 
${\cal N}=2$ SCFT operators in Chern-Simons matter theory are. 

As mentioned in \cite{AW0908}, one can study the other possibility 
where there exists a bigger $SU(3) \times U(1) \times U(1)_R$ symmetry
for the 11-dimensional lift of the same RG flow equations we discussed
in this paper. For the ${\bf CP}^2$ choice for the Einstein-Kahler
2-fold inside of $X^{p,q}$ space, the isometry is given by 
$SU(3) \times U(1) \times U(1)_R$. In the sense that this has two 
$U(1)$'s, the construction for the 4-form field strengths is similar
to each other. That is, among three $U(1)$ symmetries, only two
$U(1)$'s are preserved. Then it is nontrivial to find out the 4-forms
which should preserve these symmetries explicitly.   

There exists ${\cal N}=1$ $G_2$ critical point in 4-dimensional 
gauged supergravity. That is, this IR critical point is located at
some point in the lower plane of Figure 1.
The 11-dimensional lift of this theory, which is present in the upper plane of
Figure 1, 
was found in \cite{AI}, as mentioned in
the introduction. One can think of other 11-dimensional solution with
same RG flow equations for the $AdS_4$ supergravity fields.
Inside of 7-dimensional ellipsoid, there exists a round six-sphere
${\bf S}^6$ which has $SO(7)$ symmetry. It is an open problem 
whether one can embed the appropriate Einstein-Kahler 2-fold inside of 
${\bf S}^6$. Of course, the original 
global symmetry $G_2$ should break into a smaller group symmetry.  

\vspace{.7cm}

\centerline{\bf Acknowledgments}

I would like to thank K. Woo
for discussions. 

\appendix

\renewcommand{\thesection}{\large \bf \mbox{Appendix~}\Alph{section}}
\renewcommand{\theequation}{\Alph{section}\mbox{.}\arabic{equation}}

\section{The Ricci tensor in frame basis }

The 11-dimensional metric (\ref{11dmetric}) with (\ref{7dmetric}) 
and (\ref{delta}) 
generates the Ricci tensor
in frame basis as follows:
\bea
R_1^{\, 1} & = & -\frac{1}{9  \,\sqrt{3} \,\hat{L}^2\, u^{\frac{2}{3}} \,
  v^{\frac{4}{3}} \, (c_{\mu}^2  + u^2\, s^2_{\mu} )^{\frac{8}{3}}} 
2\left[ 2 u^8 v^2 (v^2-1)s_{\mu}^4 +
2 v^2 (v^2 +3) c_{\mu}^4 \right. \nonu \\
& + &   u^6 
\left[-2(-5+c_{2\mu})+ v^2(-11+c_{2\mu}) + 4v^4 c_{\mu}^2 \right]s_{\mu}^2
\nonu \\
&+ & \left. u^2 \left[12c_{\mu}^2 +v^2(9-13c_{2\mu})+ 
4 v^4 s_{\mu}^2\right]c_{\mu}^2
+u^4\left[6s_{2\mu}^2+v^2(5-8c_{2\mu}+5c_{4\mu}) +
v^4 s_{2\mu}^2 \right] \right]
\nonu \\
 & = & R_2^{\, 2} =R_3^{\, 3}
= -2 R_6^{\, 6} = -2 R_7^{\, 7} = -2
R_8^{\, 8}=-2 R_9^{\, 9},
\nonu \\
R_4^{\, 4} & = &
\frac{1}{18  \,\sqrt{3} \,\hat{L}^2\, u^{\frac{2}{3}} \,
  v^{\frac{10}{3}} \, (c_{\mu}^2  + u^2\, s^2_{\mu} )^{\frac{8}{3}}} 
\left[ -4 v^2 (2v^4-21v^2+27)c_{\mu}^4 -
4u^8 v^2 (2v^4-5v^2 +3) s_{\mu}^4 \right. \nonu \\
& + &   2u^2 v^2 
\left[-48+60c_{2\mu}+ v^2(15-43c_{2\mu}) + 4v^4 s_{\mu}^2 \right]c_{\mu}^2
 \nonu \\
& - &  2 u^6 \left[24c_{\mu}^2 -4v^2(7+4c_{2\mu})+
v^4 (17+ 5c_{2\mu})-4v^6 c_{\mu}^2 \right] s_{\mu}^2
\nonu \\
&- & \left. u^4 v^2 \left(33-48c_{2\mu}+27c_{4\mu} +4v^2
\left[2+4c_{2\mu}-7c_{4\mu}+v^2(-1+c_{4\mu})\right] \right) \right],  
\nonu \\
R_4^{\, 5} & = &  
\frac{1}{{6 \sqrt{3} \,\hat{L}^2\,
  v^{\frac{1}{3}}\, (c^2_{\mu} + u^2  s^2_{\mu} )^{\frac{8}{3}}}}
u^{\frac{1}{3}} 
\left(-2c_{\mu}^2(v^2-3) \right. \nonu \\
& + & \left. u^2\left[-5-11c_{2\mu}+14v^2c_{\mu}^2+
u^2\left(-11+9c_{2\mu}+
2s_{\mu}^2\left[u^2 -v^2(u^2-7)\right]\right)\right] \right) s_{2\mu}  
= R_5^{\, 4}, \nonu \\ 
R_5^{\, 5} & = & \frac{1}{18  \,\sqrt{3} \,\hat{L}^2\, u^{\frac{2}{3}} \,
  v^{\frac{4}{3}} \, (c_{\mu}^2  + u^2\, s^2_{\mu} )^{\frac{8}{3}}} 
\left[ 4u^8 v^2 (v^2-1)s_{\mu}^4 +
4v^2 (v^2+3) c_{\mu}^4 \right. \nonu \\
& + &   u^4 \left[6 - 6c_{4\mu} +v^2 
(19-16c_{2\mu}+ c_{4\mu})+ 5v^4(-1+c_{4\mu})\right] \nonu \\
& + & \left. 4 u^2 \left[6c_{\mu}^2 +v^2(3-5c_{2\mu})+
2v^4 s_{\mu}^2\right]c_{\mu}^2
+4 u^6\left[-1-v^2+v^4+(-7+5v^2 +v^4)c_{2\mu} \right] s_{\mu}^2 \right], 
\nonu \\
R_{10}^{\, 10} & = & \frac{1}{18  \,\sqrt{3} \,\hat{L}^2\, u^{\frac{2}{3}} \,
  v^{\frac{10}{3}} \, (c_{\mu}^2  + u^2\, s^2_{\mu} )^{\frac{8}{3}}} 
\left[ 4u^8 v^4 (v^2-1)s_{\mu}^4 +
4v^4 (v^2+3) c_{\mu}^4 \right. \nonu \\
& + &   u^4 v^4  \left[-2 (-11 + 8c_{2\mu}+c_{4\mu}) +5v^2 
(-1+ c_{4\mu}) \right]+ 2 u^2 v^2  c_{\mu}^2 \left[12c_{\mu}^2 +v^2 (
9-13c_{2\mu}) + 4v^4 s_{\mu}^2 \right] \nonu \\
& + & \left. 2u^6 \left[24c_{\mu}^2 -2v^2(7+13c_{2\mu})+v^4
(1+13c_{2\mu})+4v^6 c_{\mu}^2 \right] s_{\mu}^2 \right],
\nonu \\
R_{10}^{\, 11}  & = &  -\frac{2 \, u^{\frac{7}{3}} 
(v^2-1)(u^2 + 2v^2 -3)  s_{2\mu}  }
{3 \sqrt{3} \,\hat{L}^2\,
  v^{\frac{7}{3}}\, (c^2_{\mu} + u^2 \, 
s^2_{\mu} )^{\frac{5}{3}}} = R_{11}^{\,\,10},
\nonu \\
R_{11}^{\, 11} & = & \frac{1}{9  \,\sqrt{3} \,\hat{L}^2\, u^{\frac{2}{3}} \,
  v^{\frac{10}{3}} \, (c_{\mu}^2  + u^2\, s^2_{\mu} )^{\frac{8}{3}}} 
\left[ 2u^8 v^4 (v^2-1)s_{\mu}^4 +
2v^4 (v^2+3) c_{\mu}^4 \right. \nonu \\
& + &   u^2 v^2  \left[-24 c_{\mu}^2 +v^2 
(27+5c_{2\mu}) + 4 v^4  s_{\mu}^2 \right] c^2_{\mu} +u^4
v^2\left[-6s_{2\mu}^2+
2v^2(4-4c_{2\mu}+c_{4\mu}) + 7 v^4 s^2_{2\mu} \right] \nonu \\
& + & \left. u^6 \left[-24c_{\mu}^2 + 8v^2(2+5c_{2\mu})-v^4(5+29c_{2\mu})+4v^6 
c_{\mu}^2 \right] s_{\mu}^2 \right],
\label{Ricci1} 
\eea
where we introduce 
\bea
 u(r) \equiv \rho(r)^4, \qquad v(r) \equiv \cosh\chi(r). 
\nonu
\eea
We also use a simplified notation for the trigonometric function as in
$s_{\mu} \equiv \sin\mu$ and so on. By substituting the IR fixed point
values (\ref{rhochi1}) into (\ref{Ricci1}), one sees 
$R_{4}^{\,5}$ vanishes and the other components reduce to (\ref{Ricci}).

\section{The 4-form 
field strength in frame basis }

One can read off the 4-forms from (\ref{c3}), 
(\ref{a3flow}) and (\ref{geometric})
and they are given in the frame basis as follows:
\bea
F_{1234}  & = &
\frac{3^{\frac{1}{4}}\left[c_{\mu}^2(-5+\cosh2\chi)+2\rho^8(-2+
c_{2\mu}+s_{\mu}^2 \, \rho^8\, \sinh^2\chi )\right]}
{\hat{L} \,  \rho^{\frac{4}{3}}\, \cosh^{\frac{2}{3}}\chi \, 
(c_{\mu}^2+\rho^8 \, s_{\mu}^2)^{\frac{4}{3}}}, \nonu \\
F_{1235} & = & \frac{3^{\frac{1}{4}} 
\, \rho^{\frac{8}{3}}\,\left[1+ \cosh2\chi + \rho^8\, (-3+\cosh 2\chi)
\right]}
{ \hat{L} \, \cosh^{\frac{5}{3}}\chi    
\, (c_{\mu}^2+\rho^8 \, s_{\mu}^2)^{\frac{4}{3}}} \, 
s_{\mu} \, c_{\mu}, \nonu \\
F_{4568} & = & -
 \frac{3^{\frac{1}{4}} 
\, (-3+\rho^8) \, \sinh 2\chi }
{ 2 \hat{L} \, \rho^{\frac{4}{3}}\, \cosh^{\frac{5}{3}}\chi      
\, (c_{\mu}^2+\rho^8 \, s_{\mu}^2)^{\frac{1}{3}}} \, c_{\alpha+\psi} 
=-F_{4579}=
F_{469\,10}=F_{478\,10}, \nonu \\
F_{4569} & = & 
 \frac{3^{\frac{1}{4}} 
\, (-3+\rho^8)\,  \sinh 2\chi  }
{ 2 \hat{L} \,  \rho^{\frac{4}{3}}\, \cosh^{\frac{5}{3}}\chi    
\, (c_{\mu}^2+\rho^8 \, s_{\mu}^2)^{\frac{1}{3}}}\, 
s_{\alpha+\psi} =-F_{468\,10}=
F_{4578} =F_{479\,10}, \nonu \\
F_{468\,11} & = & -\frac{3^{\frac{1}{4}} \,  \rho^{\frac{8}{3}} \, 
 \mbox{sech}^{\frac{5}{3}}\chi \, 
\sinh\chi \,\left[1+ \cosh2\chi +\rho^8 \,
(-3+\cosh2\chi) \right]}
{2 \hat{L}    
\, (c_{\mu}^2+\rho^8 \, s_{\mu}^2)^{\frac{4}{3}}} \, s_{2\mu} \,
s_{\alpha+\psi}
\nonu \\
& = & -F_{479\,11}, \nonu \\
F_{469\,11} & = &  -\frac{3^{\frac{1}{4}} \,  \rho^{\frac{8}{3}} \,
\mbox{sech}^{\frac{5}{3}}\chi \,
\, \sinh\chi \,\left[1+ \cosh2\chi +\rho^8\,
(-3+\cosh2\chi)\right]}
{2 \hat{L}    
\, (c_{\mu}^2+\rho^8 \, s_{\mu}^2)^{\frac{4}{3}}} \,  s_{2\mu} \, c_{\alpha+\psi}
\nonu \\
&=& F_{478\,11}, \nonu \\
F_{568\,10}  & = &  \frac{3^{\frac{1}{4}} 
\, \rho^{\frac{8}{3}}\,(-1+ \rho^8) \,\sinh\chi}
{\hat{L} \, \mbox{sech}^{\frac{1}{3}}\chi   
\, (c_{\mu}^2+\rho^8 \, s_{\mu}^2)^{\frac{4}{3}}}\,s_{2\mu} \, 
s_{\alpha+\psi} =-F_{579\,10}, 
\nonu \\
F_{568\,11} & = & 
 - \frac{3^{\frac{1}{4}} 
\,  (3 + \rho^8)\, \sinh\chi}
{\hat{L} \,  \rho^{\frac{4}{3}}  \, \cosh^{\frac{2}{3}}\chi \, 
(c_{\mu}^2+\rho^8 \, s_{\mu}^2)^{\frac{1}{3}}}\,s_{\alpha+\psi}
=-F_{579\,11}=F_{69\,10\,11}=F_{78\,10\,11}, 
\nonu \\
F_{569\,10}  & = &   \frac{3^{\frac{1}{4}}\,
\rho^{\frac{8}{3}} \, (-1 + \rho^8)  \, \sinh\chi }
{\hat{L} \, \mbox{sech}^{\frac{1}{3}}\chi    
\, (c_{\mu}^2+\rho^8 \, s_{\mu}^2)^{\frac{4}{3}}}  \, c_{\alpha+\psi}
\, s_{2\mu} 
=F_{578\,10},
\nonu \\
F_{569\,11} & = &  - \frac{3^{\frac{1}{4}}
\, (3 + \rho^8) \, \sinh\chi }
{\hat{L} \, \rho^{\frac{4}{3}} \, \cosh^{\frac{2}{3}}\chi \,     
(c_{\mu}^2+\rho^8 \, s_{\mu}^2)^{\frac{1}{3}}}  \, c_{\alpha+\psi}
=F_{578\,11}=-F_{68\,10\,11}=F_{79\,10\,11}. 
\label{F41}
\eea
For simplicity, we ignored the $r$ dependence on $\rho$ and $\chi$ in
the right hand side of (\ref{F41}).
When we substitute the UV fixed point value (\ref{rhochi}) into
(\ref{F41}), then only $F_{1234}$ is nonzero.
When we substitute the IR fixed point
values (\ref{rhochi1}) into (\ref{F41}), one sees 
$F_{\mu\nu\rho 5}, F_{4mnp}$ and $F_{45mn}$
where $\mu, \nu, \rho =1, 2, 3$ and $m,n, p =6, 7, \cdots, 11$
vanish due to the $\sinh\chi(r)$ 
and the other components reduce to (\ref{F4}).
These vanishing 4-forms have either $1+ \cosh2\chi + \rho^8\,
(-3+\cosh 2\chi)$, which leads to $(3-\rho^8)$ for the condition $\cosh2\chi=2$, 
or $(-3+\rho^8)$. 
This nontrivial boundary conditions also occur for the ${\cal N}=1$
$G_2$ 
M2-brane flow in 11-dimensions \cite{AI}.


\end{document}